\documentclass{article}
\usepackage{ijcai22}

\usepackage{times}
\usepackage{soul}
\usepackage{url}
\usepackage[hidelinks]{hyperref}
\usepackage[small]{caption}
\usepackage{graphicx}
\usepackage{amsthm}
\usepackage{booktabs}
\usepackage{algorithm}
\usepackage{algorithmic}
\usepackage{amsmath,amssymb,fancyhdr,wasysym,graphicx}
\usepackage{graphicx} 
\usepackage{multirow} 
\usepackage{amsfonts}
\usepackage{url}
\usepackage{xcolor}
\usepackage{threeparttable}
\usepackage{makecell}
\usepackage{subfigure}
\usepackage{enumitem}
\usepackage{fancyhdr}%
\usepackage{natbib}
\title{Imperceptible Sample-Specific Backdoor to DNN with Denoising Autoencoder}

\author{
Xiangqi Wang$^1$
\and
Mingfu Xue$^2$\and
Kewei Chen$^{3}$\and
Jing Xu$^4$\and
Wenmao Liu$^5$\and
Leo Yu Zhang$^6$\and
Yushu Zhang$^3$
\affiliations
$^1$School of Mathematics and Statistics, Hunan First normal University, Changsha, China\\
$^2$School of Communication and Electronic Engineering, East China Normal University, Shanghai, China\\
$^3$College of Computer Science and Technology, Nanjing University of Aeronautics and Astronautics, Nanjing, China\\
$^4$College of Computer Science and Electronic Engineering, Hunan University, China\\
$^5$Innovation Center, NSFOCUS Information Technology Co., Ltd., Beijing, China\\
$^6$School of Information and Communication Technology, Griffith University, QLD, Australia
\emails
\ xiangqi.wang@foxmail.com, mfxue@cee.ecnu.edu.cn, chenkewei@nuaa.edu.cn, xj123@hnu.edu.cn, liuwenmao@nsfocus.com, leo.zhang@griffith.edu.au, yushu@nuaa.edu.cn}

\begin{document}

\maketitle

\thispagestyle{fancy} %

\lhead{} %
\chead{} %
\rhead{} %
\lfoot{} %
\cfoot{\thepage} %
\rfoot{}
\renewcommand{\headrulewidth}{0pt} %
\renewcommand{\footrulewidth}{0pt}
\pagestyle{fancy}
\cfoot{\thepage}

\begin{abstract}
The backdoor attack poses a new security threat to deep neural networks.
Existing backdoor often relies on visible universal trigger to make the backdoored model malfunction, which are not only usually visually suspicious to human but also catchable by mainstream countermeasures.
We propose an imperceptible sample-specific backdoor that the trigger varies from sample to sample and invisible. Our trigger generation is automated through a desnoising autoencoder that is fed with delicate but pervasive features (i.e., edge patterns per images).
We extensively experiment our backdoor attack on ImageNet and MS-Celeb-1M, which demonstrates stable and nearly 100\% (i.e., 99.8\%) attack success rate with negligible impact on the clean data accuracy of the infected model. The denoising autoeconder based trigger generator is reusable or transferable across tasks (e.g., from ImageNet to MS-Celeb-1M), whilst the trigger has high exclusiveness (i.e., a trigger generated for one sample is not applicable to another sample). Besides, our proposed backdoored model has achieved high evasiveness against mainstream backdoor defenses such as Neural Cleanse, STRIP, SentiNet and Fine-Pruning.
\end{abstract}

\section{Introduction}
Deep neural network (DNN) has been successfully applied in many fields, such as image classification, speech recognition and natural language processing. Although DNN has excellent performance in these fields, researchers found that in the entire model life cycle from model training to inference, it is vulnerable to various attacks~\citep{ref3,ref2}. In the model prediction stage, adversarial example attack adds subtle perturbations to the input to fool the model. In the model training stage, poisoning attack modifies the training set to affect the model function, which can insert backdoor to the model.

Backdoor attack is one of the most insidious attack occurred in the DNN training phase~\citep{ref1,gao2020backdoor}. The attacker only needs to poison a small fraction of training samples to inject backdoors into the model. For samples with triggers added, the backdoored model will output the target label pre-specified by the attacker.
The backdoored model functions normally in predicting benign samples in the absence of the trigger.
Therefore, backdoor attack can be immune to some effective defense methods against general poisoning attacks that aim to degrade the overall accuracy performance of the infected model. For example, poisoning attack but not backdoor attack can be detected by model performance on a hold-out set. Thus, the trigger mode of backdoor attack is more insidious than poisoning attack.

\begin{figure}[t]
	\setlength{\abovecaptionskip}{0.1cm}
	\centering
	\includegraphics[width=\linewidth]{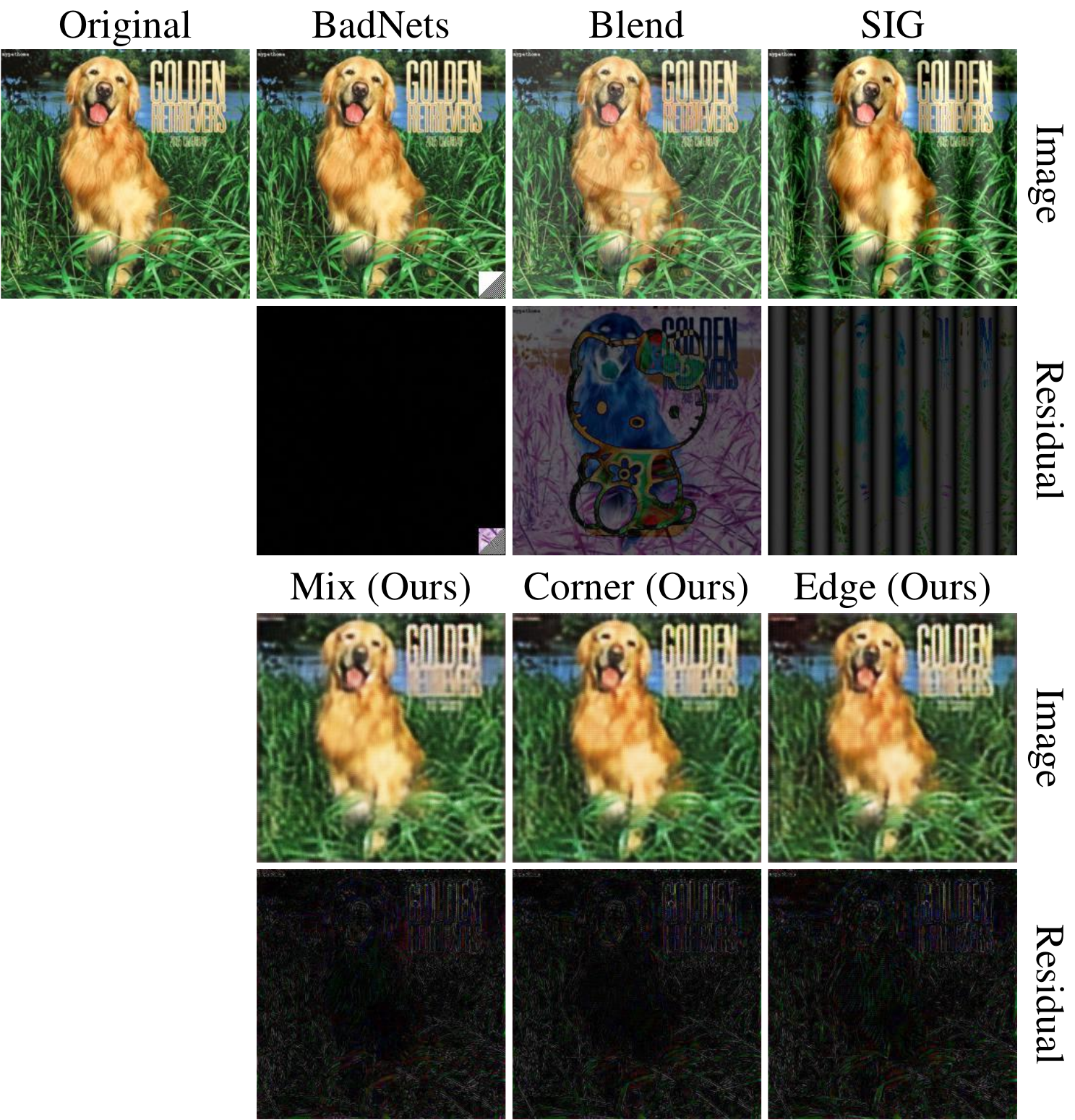}
	\caption{Comparison between backdoor examples generated by our method and existing backdoor attacks.}
	\label{fig1}
\end{figure}

Since backdoor attack was introduced into DNN, many attack methods~\citep{r2,blend,wanet,sig} have been proposed, but these methods suffer from two main limitations: the trigger pattern of the backdoor is fixed or universal and the human eye can recognize the trigger presented in the abnormal image. The trigger mode is exemplified in Figure~\ref{fig1}. Therefore, many defense methods~\citep{r6,r9,nc} are designed against universal triggers. They are effective to capture universal trigger backdoor due to the same and strong universal pattern. For the visibility of triggers, when the data curator checks the training set, the suspected samples can be identified and eliminated.

Recently, few advanced backdoor trigger crafting methods have been proposed\citep{r4,ISSBA}. In their approach, the generation pattern of backdoor triggers obviates the shortcomings of conventional trigger generation methods. More specifically, triggers are dynamic and change for different inputs. Such sample-specific triggers can trivially evade most mainstream backdoor detections that rely on the assumption of universal trigger patterns.
Nonetheless, these works still have deficiencies: triggers are not often exclusive~\citep{r4}, or the trigger generator is not transferable across specific trained models and tasks~\citep{ISSBA}.

To resolve the shortcomings of the above backdoor attack methods, this work proposes a backdoor attack whose triggers vary with input and are imperceptible to human eyes, which can thus resist the mainstream backdoor defense methods~\citep{nc,fp,strip,sentinet}. At the same time, as shown in Table \ref{t1}, compared with the existing dynamic backdoor attacks, the triggers generated by our method exhibit high exclusiveness, and the trigger generator is transferable.

Generally, we first inject a sample-specific feature (i.e., visible edge pattern) into an incoming image and then transform this feature into an imperceptible backdoor trigger through the denoising autoencoder, whilst altering the label to the target label. Such transformed images will be regarded as poisoned samples, which will be mixed with benign samples at the training phase to train the victim model. In contrast to~\citep{r4} where a trigger generator has to be trained jointly with a targeted model, our attack does not require any knowledge of the victim model (i.e., architecture, training loss). Our attack only needs to poison a small fraction of training samples provided to the data curator, which does not need to control or have knowledge of model training (e.g., model loss function and model structure) performed by the victim user. As exemplified in Figure~\ref{fig1}, in addition to sample-specific, the trigger of our attack is imperceptible, which cannot be visually recognized.


\begin{table}[t]
    \centering
    \scalebox{0.65}{
    \begin{tabular}{ccccc}
        \toprule
             \textbf{Work}  &
             \textbf{\makecell[c]{No \\ training \\ control}} &
             \textbf{Imperceptibility}& \textbf{Transferability}  & \textbf{Exclusivity}  \\
        \midrule
             \cite{r4}& $\times$ & $\times$ & $\times$ & \checkmark \\
             \cite{ISSBA}& \checkmark & \checkmark & \checkmark & $\times$\\
            Ours & \checkmark & \checkmark & \checkmark & \checkmark \\
        \bottomrule
    \end{tabular}}
    \caption{A comparison of dynamic (sample-specific) backdoor attacks}
    \label{t1}
\end{table}

\vspace{0.05cm}
\noindent{\bf Our Contributions.} Main contributions of this work are threefold:

\begin{itemize}[leftmargin=*]
    \item We propose a backdoor attack framework based on denoising autoencoder, which generates trigger samples by denoising the injected (perceptible) image-specific feature. Therefore, the generated triggers are imperceptible and sample-specific. A small fraction (i.e., 1\%) of trigger samples is able to successfully implant the backdoor without any prior knowledge or control of the victim model.

    \item We propose diverse feature injection means to be compatible with the denoising autoencoder. The triggers generated by our method are also exclusive. The trigger generator (i.e., denoising autoencoder) trained for a given model of a given task is transferable to not only other models of the same task but also to other tasks.


    \item We validate the effeciency of our method on large-scaled ImageNet and MS-Celeb-1M datasets. On the premise of not deteriorating the accuracy of the model, our method can achieve 99.8\% attack success rate. At the same time, our method bypasses a number of diverse mainstream backdoor defenses of Neural Cleanse, STRIP, SentiNet and Fine-Pruning.

\end{itemize}

\section{Related Work}
\subsection{Backdoor Attacks}
Backdoor is a newly revealed security threat that injects hidden malicious functions into the artificial intelligent system.
Recently, a large number of studies have shown that diverse DNNs are vulnerable to backdoor attacks. According to the characteristics of triggers, existing backdoor attacks can be generally divided into two categories: (1) universal (sample-agnostic) backdoor; (2) dynamic (sample-specific) backdoor.

\noindent{\bf Universal (sample-agnostic) backdoor.}
A trigger pattern stamped on different samples is fixed, which is easy to be captured by mainstream backdoor defenses.
\citet{badnet} proposed BadNets, which works in the way of data poisoning attack. In the model training stage, the attacker modifies small partial training set by stamping the universal trigger into samples, and corresponding changing the sample label to the target label.
In the prediction stage, the input of the added trigger can activate the backdoor.
After that, \cite{r2} proposed a Trojan attack, which generates triggers by manipulating specific neurons in the clean model, and then fine-tunes the clean model with poisoned data to inject backdoor.
\cite{blend} blended the original image and trigger to get poisoned images, which is different from the patch injection method in BadNets. Poisoned images generated by this method have a certain degree of concealment, but the concealment decreases with the increase of the image blending ratio.
\cite{sig} proposed a clean-label backdoor attack with a sinusoidal signal (SIG) as a trigger. However, more perturbation is required for high-resolution images, which significantly degrades image quality.
The triggers mentioned above are also belong to the visible backdoor attacks, and humans can easily identify abnormal trigger patterns.

\noindent{\bf Dynamic (sample-specific) backdoor. }
In contrast to the universal backdoor above, dynamic backdoor has a flexible trigger pattern, making its generated triggers changeable upon input samples.
\cite{r4} proposed a perceptible
but dynamic trigger generation mode. During the training of a backdoored classification model, they jointly train a trigger generation network, which can generate different triggers for different inputs. However, the trigger generation network is bounded to a specific backdoored model, each backdoored model needs a different trigger generation network, which reduces its real threat due to lack of e.g., transferability to different models.
\cite{ISSBA} proposed to embed the trigger into the DNN through image steganography, making it difficult to distinguish the poisonous image from the benign image visually. The triggers generated by their method are not always sample-specific. That is, a trigger of a sample can still be used to activate a different sample when it is stamped, exhibiting low exclusiveness.
Such a low exclusiveness is obviated by our attack.
\cite{wanet} defined the sampling position of the reverse warping for each point in the target image through the image warping function, thereby generating a covert backdoor image. However, in order to bypass the mainstream defense methods, this method needs to poison 30\% of the training set. In their attack methods, in addition to poisoning 10\% of the data, they also need 20\% of the noise data to bypass the defense methods. Such a high attack budget is hard to be achieved in practice, especially in the data outsourcing scenario. Our attack requires only 1\% poison rate to succeed.
In the clean-label backdoor attack proposed by \cite{r11}, the poisoned image is made close to the image of the target class in pixel space by a constraint function, while close to the source image patched with triggers in feature space. This attack method requires a strong assumption: the attacker has knowledge of the victim model structure or even weights~\citep{ma2022macab}.
In this paper, our proposed imperceptible backdoor attack does not require the attacker to have any knowledge of the training process including the used model structure.

\subsection{Backdoor Defenses}
With an in-depth study of backdoor attack methods, various backdoor defenses have been proposed. According to different defensive objectives, backdoor defenses can be generally divided into two categories: model-based defense and data-based defense~\citep{gao2020backdoor}.

\noindent{\bf Model-based Defenses. } Model-based defenses aim to identify backdoored models, and then remove backdoors or avoid its afterward usage. In the Fine-pruning proposed by \cite{fp}, the pruning and fine tuning of the model are combined. They first prune the dormant neurons in the model when benign samples are inputs, and then fine tune the model to reduce the degrade of clean data accuracy.
\cite{nc} detected whether the model was implanted a backdoor by reversing the trigger and identifying the attacker target label, and then weakened the backdoor success rate through the usage of the reversed trigger to perform backdoor unlearning.
\cite{abs} scanned each neuron in the model and directly modify the activation value, observe the difference of the model output to find the backdoor neuron and reverse the backdoor trigger.

\noindent{\bf Data-based Defenses. } Data-based defenses aim to identify abnormal data and eliminate the possibility of backdoor implantation and activation at the data level. \cite{strip} proposed the STRIP, they found that backdoor samples were insensitive to strong perturbations while benign samples were sensitive. Based on this characteristic, they detected trigger samples in the testing phase of the model---the STRIP can also be used during offline when the training set is accessible. \cite{sentinet} also performed backdoor detection on the samples, they adopted a saliency map (Grad-CAM~\citep{gradcam1}) to locate trigger regions and filter out trigger samples.

Existing defense methods are generally designed for universal triggers, and thus fail for dynamic triggers. Moreover, the size of the trigger also affects the defense effect (e.g.,~\citep{nc,sentinet}). The computational cost of some defense methods is proportional to the number of labels, and when the number of labels is large, the computational cost will be high (e.g.,~\citep{nc}). In our attack, the generated triggers are dynamic and cover the whole image, so they are evasive against mainstream defenses.

\section{Imperceptible Sample-Specific Trigger with Denoising Autoencoder}\label{OVERVIEW}

In this section, we first introduce the pipeline of a backdoor attack, and then define the threat model, followed by an elaboration of our proposed attack.

\subsection{Pipeline of Backdoor Attacks}
For an image-classification task, a DNN model $f(\cdot)$ maps a feature vector $x$ from input image space $X$ to a label $y$ from output category space $Y$, that is, $f: x \mapsto y (x \in X, y \in Y)$. If a backdoor is injected into the model, for a benign input $ x_c $ with no trigger, the model outputs a correct classification result  $y_c=f(x_c) $.  For a poisoned input $x_p $ with a trigger carried, the trigger will activate the backdoor, making the model output an attacker-targeted label $y_t\in Y $.

In the backdoor attack, the attacker modifies part of the training set to achieve the purpose of injecting the backdoor, that is, the benign sample $(x,y)$ is modified by the trigger generation mode $p(\cdot)$ to gain a trigger sample and the target label generation mode $h(\cdot)$ to obtain poisoned samples $(x_p, y_t)=(p(x), h(y)) \in D_{poison}$. The training set is now a mixture of clean and poisonous samples, which can be expressed as:
\begin{equation}
D_{train}= D_{clean}\cup D_{poison},
\end{equation}

where the poisoning ratio $\rho$ is the ratio of the number of elements in the poisoned dataset $D_{poison}$ to the training set $D_{train}$, which is expressed as:
\begin{equation}
\rho=\frac{|D_{poison}|}{|D_{train} |}.
\end{equation}

\subsection{Threat Model}

\noindent{\bf Knowledge and Capability.} We assume that the attacker can only access and tamper with part of the model training data, but cannot obtain any model information, nor participate in the model training process.
In addition, in the model inference stage, the attacker can only obtain the inference result $y$ of the model for the input $x$, and cannot manipulate the inference process of the model. Moreover, the attacker should reduce its attack budget, e.g., using a small poison rate.

\noindent{\bf Attacker Goals.} The attacker goals are as follows:

\begin{itemize}
    \item The backdoored model should have high probability of classifying a trigger image into an attacker-targeted label;

    \item For benign input, the classification accuracy of the backdoored model should be close to its clean model counterpart. That is, the model needs to behave normally when the trigger is not present;

    \item The content of a trigger image should be almost the same to its clean counterpart to evade human inspection in the training and inference stage;

    \item The backdoor attack is evasive against backdoor defenses as much as possible.
\end{itemize}

\subsection{Our Proposed Attack}
In order to achieve the attacker goals, we propose imperceptible sample-specific trigger backdoor attack based on denoising autoencoder. Specifically, the attack process is divided into two steps:

1) Training a denoising autoencoder to convert specific injected features (i.e., edge pattern, corner points) into imperceptible triggers;

2) According to the attack budge---poisoning ratio $\rho$, poison corresponding fractional clean images with imperceptible sample-specific triggers assisted by the above trained denoising autoencoder and modify their labels as the target label.

Then the victim user may train the DNN according to the standard procedure using the curated data that has the poisoned images. The poisoned image content generated by the denoising autoencoder is visually similar to the original image, but retains part of the injected feature information. The human eye is insensitive to this injected feature, but DNNs can capture it during the training process and respond to this backdoor feature during the inference stage.

\begin{figure}[t]
	\setlength{\abovecaptionskip}{0.1cm}
	\centering
 \scalebox{0.75}{
\includegraphics[width=\linewidth]{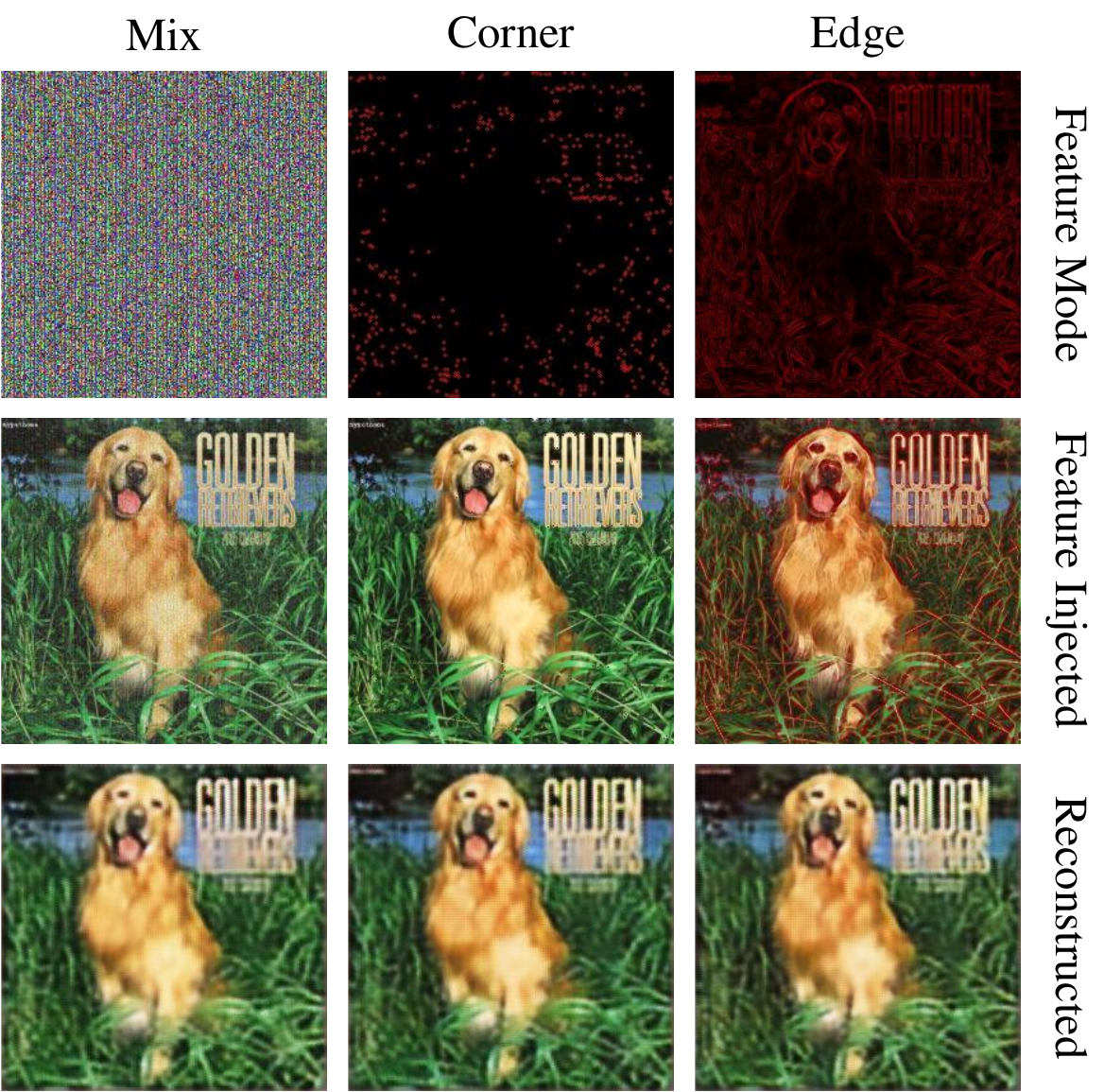}}
	\caption{Examples of three feature injection modes and images after feature injection and denoising autoencoder reconstruction.}
	\label{figfi}
\end{figure}

\vspace{0.05cm}
\noindent{\bf Feature Injection Mode. }
As shown in Figure \ref{figfi}, this work considers three modes of feature injection:
\begin{itemize}
    \item Image mixing, which mixes a clean image with a noise image in a certain proportion;

    \item Corner points, which extract corner points of a clean image by an algorithm (e.g., Harris~\cite{harris} and Shi-Tomasi~\cite{goodfeature}) and fill them with a specific color;

    \item Edge, which extracts the edge of a clean image and fills them with a specific color using an edge detection algorithm (e.g., Sobel~\cite{sobel}).
\end{itemize}

Though we specifically use three modes of feature injection in this work, the attacker can in fact select other feature injection modes through snowflake noise, streak noise, etc, following the same framework.
However, for the sake of ensuring the denoising autoencoder works properly, there are preferable criteria.
Firstly, the coverage of image modification when injecting features is preferred to be the entire image, because the denoising autoencoder will reconstruct the entire image. Secondly, when injecting features, the semantic information of the image should not be modified, such as patched a sticker.

\begin{figure}[t]
\setlength{\abovecaptionskip}{0.1cm}
	\centering
    \scalebox{0.9}{\includegraphics[width=\linewidth]{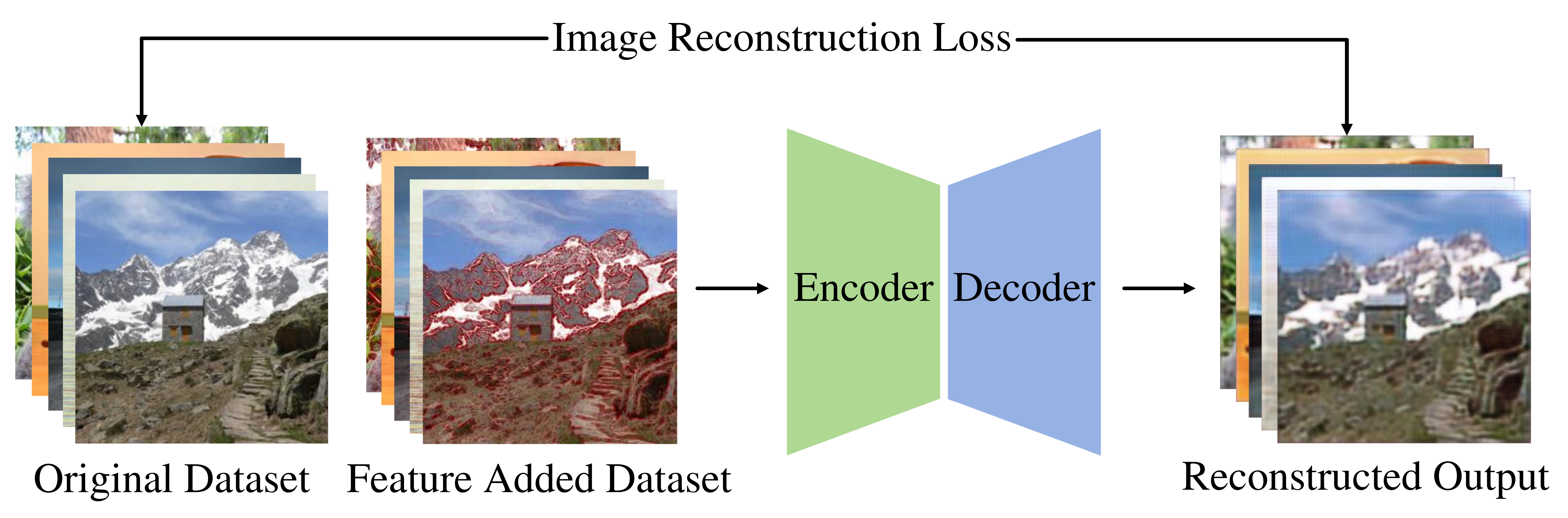}}
	\caption{The training process of the denoising autoencoder.}
	\label{fig2}
\end{figure}

\vspace{0.05cm}
\noindent{\bf Training of Denoising Autoencoder.} We denote the denoising autoencoder as $E(\cdot)$, the feature injection mode as $I(\cdot)$, then the process of generating a trigger image can be expressed as $E(I(x))$. We record the data used to train the denoising autoencoder that does not overlap with the model training set as $D_{benign}$, then the training set of denoising autoencoder $E$ is denoted as $ \{(I(x),x) | x \in D_{benign}\}$. The training process of the denoising autoencoder is shown in Figure \ref{fig2}. In order to meet the third attacker goal, the distance between the reconstructed image $E(I(x))$ and the original image $x$ should be as small as possible, so we use the $L_2$ norm as the loss function:
\begin{equation}
L(I(x),x)=\frac{1}{m}\sum_{i=0}^n (I(x)-x)^2.
\end{equation}

\noindent{\bf Poisoned Dataset Generation.} Our proposed method for generating covert trigger-carrying image is as follows. First, a specific feature is injected into a clean image, which feature is then converted into invisible trigger of the image
through the denoising autoencoder. The attacker modifies the label of poisoned images to $y_t\in Y$ to obtain the poisoned trainging dataset expressed as:
\begin{equation}
  D_{poison}=\{(x_p,y_t)|x_p=E(I(x)),(x,y) \in D_{benign}\}.
\end{equation}

Examples of the poisoned images are shown in the third row in Figure~\ref{figfi}. Triggers in poisoned images are imperceptible to the human eye.

\begin{figure}[t]
	\setlength{\abovecaptionskip}{0.1cm}
	\centering
	\scalebox{0.9}{\includegraphics[width=\linewidth]{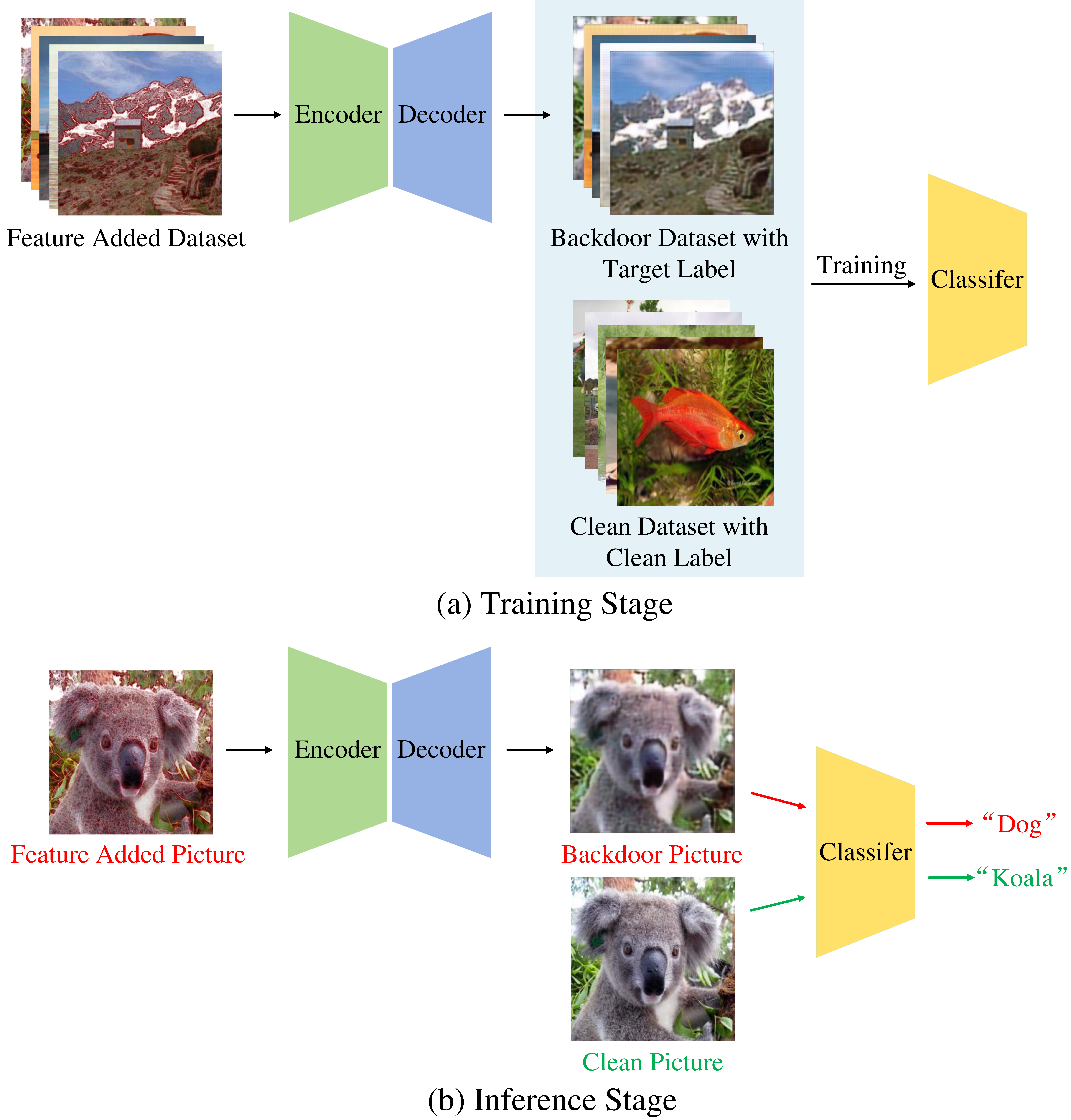}}
	\caption{Attack pipeline.}
	\label{fig3}
\end{figure}

\vspace{0.05cm}
\noindent{\bf Attack Pipeline.} The backdoor attack pipeline is shown in Figure \ref{fig3}. In the model training stage, the attacker generates a poisoned dataset, which is mixed with the clean dataset by the victim user as the training set $D_{train}$. The victim trains the model, which is inserted in a backdoor.
During the inference stage, the backdoored model gives correct classification results for benign inputs, while for input containing the trigger, its classification result will be the target label.

\section{Experimental Analysis}\label{Experimental Analysis}

\subsection{\bf Experimental Setup}

\noindent{\bf Dataset.} We consider two popular classification tasks, general image classification and face recognition. For the first task, we conduct experiments on the ImageNet~\cite{ImageNet} dataset. We randomly selected a subset of 200 classes with 125,000 images, and the split ratio of training and testing sets is set to 8:2. The image size is $256\times 256 \times 3$. For the second task, we conduct experiments on the MS-Celeb-1M~\cite{ms1m} dataset. We randomly selected a subset of 155 classes, with 43,400 images for training and 10,850 images for testing. The image size is $112\times 112 \times 3$.

\vspace{0.05cm}
\noindent{\bf DNN Structure.} For the DNN structure used by the victim user, we consider two popular networks: ResNet-18~\cite{res18} and VGG-16~\cite{vgg16}, which are used as training models for image classification and face recognition tasks, respectively.

\vspace{0.05cm}
\noindent{\bf Evaluation Metric.} We use Attack Success Rate (ASR) and Clean Data Accuracy (CDA) to evaluate backdoor attack performance. Specifically, ASR is defined as the ratio between the successfully attacked poisoned samples and the total poisoned samples. CDA is the performance on the clean test set (i.e., the accuracy on non-trigger carrying test images). The "$\times$" represents the percentage of the backdoor model's CDA relative to its clean model counterpart. In addition, we use Mean Squared Error (MSE)\cite{mse}, Peak Signal-to-Noise Ratio (PSNR)\cite{psnr}and Structure Metric (SSIM)\cite{ssim} to evaluate the stealthiness of the poisoned images.

\vspace{0.05cm}
\noindent{\bf Default Settings for Attacking.} When we inject features through noise mixing mode, we randomly generate a noisy image and blend it to clean images at a ratio of 20\%. When we inject features through corner points, we select Shi-Tomas corner detection algorithm to extract corner points of an image and modify their colors (R: 139, G: 0, B: 0). When we inject features through edge pattern, we select the Sobel operator to extract edges of an image and modify their colors (R: 139, G: 0, B: 0). Figure~\ref{figfi} shows examples of images after feature injection (i.e., artifacts can be visible) and denoising autoencoder reconstruction (i.e., artifacts are imperceptible). We set the poisoning ratio $\rho=10\%$ in all attacks (i.e., small poison ration is studied later).
In the attack comparison experiment, the $\rho$ of BadNets and Blend is 10\%. Since SIG only needs to poison the data of the target class, we poisoned up to 70\% of the target class data\footnote{We have tried lower poison rates for this SIG attack, but resulted in a quite low ASR.}, despite the $\rho$ to the entire dataset is 0.35\% in ImageNet and 0.45\% in MS-Celeb-1M. The target label is $y_t=0$ in all attacks.

We use the \texttt{SGD} optimizer with a momentum of 0.9 for training classifier $f$. For ImageNet, we set the training batch to 128, the epoch number is 70 and the learning rate is initially 0.001, which is further set as 0.0001 and 0.00001 at epoch 15 and epoch 35, respectively. For MS-Celeb-1M, we set the training batch to 32, the epoch number is 70, and the learning rate is 0.001. We use the \texttt{Adam} optimizer for training the denoising autoencoder, and set the batch size to 32, the epoch number is set to 300, and the learning rate is set to 0.0003.

\vspace{0.05cm}
\noindent{\bf Machine Configuration.} All experiments were conducted on a server equipped with an Intel Xeon(R) Cold 5218R 2.1GHz processor, a 64GB RAM and a NVIDIA GeForce RTX 3080 GPU.

\subsection{\bf Attack Performance}
In this section, we evaluate backdoor attacks based on denoising autoencoders in terms of both attack effectiveness and attack concealment, and compare it with the typical visible trigger attack of BadNets~\citep{badnet} and invisible trigger attacks of Blend~\citep{blend} and SIG~\citep{sig}.

\vspace{0.05cm}
\noindent{\bf Attack Effectiveness.} We conduct backdoor attack experiments on the ResNet-18 model trained on the ImageNet dataset and the VGG-16 model trained on the MS-Celeb-1M dataset. The experimental results of the effectiveness comparison of different attack methods are shown in Table \ref{table1}. All three of our feature injection modes can achieve up to 99.8\% ASR with negligible impact on CDA of the backdoored model. In most cases, our methods have higher ASR and smaller CDA than BadNet, Blend, and SIG. In Section \ref{Ablation Studies}, we further discuss the effect of reducing the poisoning ratio on the CDA and ASR. The results show that when the poisoning ratio is 1\%, the ASR is still as high as 94\% on the ImageNet dataset and higher than 99\% on the MS-Celeb-1M dataset, while the CDA degradation is less than 1\%. This affirms that our method can complete an effective backdoor attack with a negligible CDA cost.

\begin{table}[t]
\centering
\scalebox{0.8}{
\begin{tabular}{c|cc|cc}
    \toprule %
    \multirow{2}*{\textbf{Methods}}&
    \multicolumn{2}{c|}{\textbf{ImageNet}}&
    \multicolumn{2}{c}{\textbf{MS-Celeb-1M}}\\
           \cline{2-5}   %
           ~& \textbf{CDA} & \textbf{ASR} & \textbf{CDA} & \textbf{ASR} \\
    \midrule %
Benchmark & 86.26 & -  & 98.30 & - \\
BadNets & 85.47 (99.08$\times$) & 99.71  & 97.96 (99.65$\times$)& \textbf{100.00} \\
Blend & 85.46 (99.07$\times$) & 98.27  & 97.97 (99.66$\times$)& 99.97 \\
SIG & 84.77 (98.27$\times$) & 77.73 & 97.04 (98.72$\times$) & 74.40 \\
Mix(ours) & \textbf{85.94 (99.63$\times$)} & \textbf{99.87}  & 98.13 (99.83$\times$)& 99.94 \\
Corner(ours) & 85.92 (99.61$\times$) &  99.83 & \textbf{98.14 (99.84$\times$)}& 99.94 \\
Edge(ours) & 85.78 (99.44$\times$) & 99.83 & 98.08 (99.78$\times$) & 99.98 \\
    \bottomrule %
\end{tabular}}
\caption{Comparison of different attack effects (CDA/ASR)}
\label{table1}
\end{table}

\vspace{0.05cm}
\noindent{\bf Attack Concealment.} Table \ref{tableconcealment} presents the MSE, PSNR and SSIM performance of different backdoor attacks. All our three feature injection modes outperform BadNets, Blend, and SIG on both MSE and PSNR on both datasets. The SSIM of our method outperforms Blend and SIG on the MS-Celeb-1M dataset and SIG on the ImageNet dataset. The reason why our method scores worse than BadNets on PSNR is that BadNets only modifies the color of \textit{few pixels} in a large-sized image, but the trigger of BadNets is unnatural and easier to be recognized by human eyes.
Overall, the results affirm that the triggers generated by our method are stealthier.

\begin{table}[t]
\centering

\begin{threeparttable}\scalebox{0.75}{
     \begin{tabular}{c|ccc|ccc}
        \toprule %
        \multirow{2}*{\textbf{Methods}}&
        \multicolumn{3}{c|}{\textbf{ImageNet}}&
        \multicolumn{3}{c}{\textbf{MS-Celeb-1M}}\\
               \cline{2-7}   %
               ~& \textbf{MSE$\downarrow$} & \textbf{PSNR$\uparrow$} & \textbf{SSIM$\uparrow$} & \textbf{MSE$\downarrow$} & \textbf{PSNR$\uparrow$} & \textbf{SSIM$\uparrow$}  \\
        \midrule %

    BadNets & 268.58 & 24.14 & \textbf{0.98}\tnote{*} & 301.45 & 23.73 & \textbf{0.98}\tnote{*}\\
    Blend & 468.13 & 21.69 & 0.84 & 466.61 & 21.63 & 0.79\\
    SIG & 714.97 & 19.55 & 0.71  & 714.97 & 19.60 & 0.53\\
    Mix(ours) & 245.86 & 24.82 & 0.73 & 175.84 & 26.04 & 0.82 \\
    Corner(ours) & \textbf{233.28} & \textbf{25.06} & 0.74 & \textbf{160.84} & \textbf{26.41} & 0.83 \\
    Edge(ours) & 263.78 & 24.50 & 0.73 & 181.91 & 25.93 & 0.82 \\
        \bottomrule %
        \end{tabular}}
 \begin{tablenotes}
    \footnotesize
    \item[*] The reason SSIM is high on BadNets is that the trigger only\\modifies 1\% of the pixels in the original image.
\end{tablenotes}
\caption{Trigger concealment comparison of different backdoor attacks}
\label{tableconcealment}
\end{threeparttable}
\end{table}

\vspace{0.05cm}
\noindent{\bf Time Overhead.}
We measured the average time to train the denoising autoencoder and the time to generate poisoned data.
The time to train the denoising autoencoder is less than 3 minutes in our experiments, which is efficient in the offline phase. The time to produce an imperceptible sample-specific trigger image through the trained denoising autoencoder is no more than 26ms, which can be done in real-time during the online attack phase.

\subsection{\bf Evasiveness Against Backdoor Defenses}

In this part, we evaluate the ability of the backdoored model to resist several mainstream defense technologies. For model-based defense methods, we evaluate Fine-Pruning~\citep{fp} and Neural Cleanse~\citep{nc}. For data-based defense methods, we evaluate STRIP~\citep{strip} and SentiNet~\citep{sentinet}.

\begin{figure}[t]
\setlength{\abovecaptionskip}{0.1cm}
	\centering
	 \scalebox{0.85}{\includegraphics[width=\linewidth]{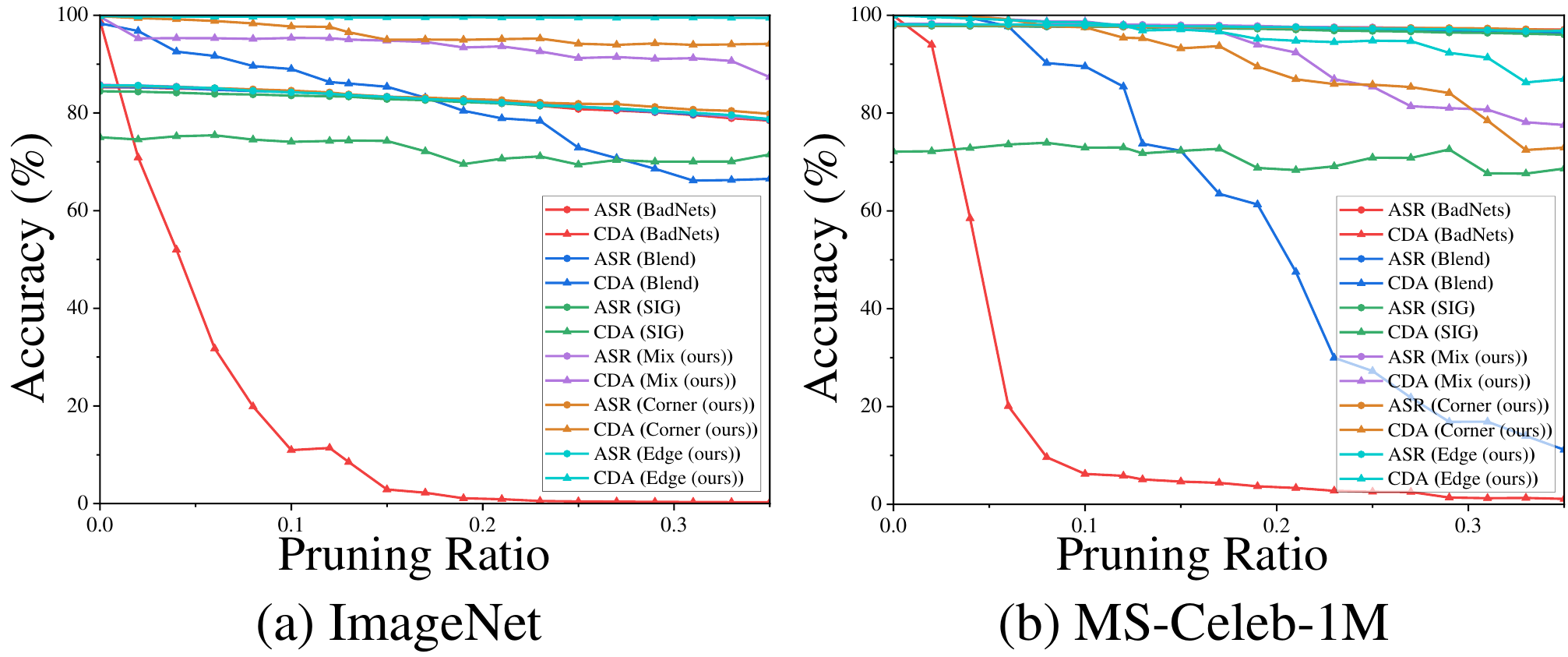}}
	\caption{The effect of different attack methods against Fine-Pruning defense method.}
	\label{figFP}
\end{figure}

\vspace{0.05cm}
\noindent{\bf Fine-Pruning.} In the process of model prediction, clean samples, and trigger samples will activate different neurons, Fine-Pruning records the average activation of each neuron under clean samples, and iteratively prunes dormant DNN neurons in order to remove compromised neurons. To remedy CDA drop, a later fine-tuning is taken through held-out clean data.
We compare our method with BadNets, Blend, and SIG attacks, which results are shown in Figure \ref{figFP}. When the pruning ratio reaches 30\%, BadNets fail on both ImageNet and MS-Celeb-1M datasets. The ASR of Blend is reduced by 32\% and 83\% respectively, and the ASR of SIG is not more than 70\%, while the ASR of our method is only reduced by 4.82\% on the ImageNet dataset on average, and the lowest is still 91.08\%. On the MS-Celeb-1M dataset, the ASR is reduced by 14.15\% on average, and the lowest is still up to 81.01\%.

\begin{figure}[t]
\setlength{\abovecaptionskip}{0.1cm}
	\centering
	 \scalebox{0.85}{\includegraphics[width=\linewidth]{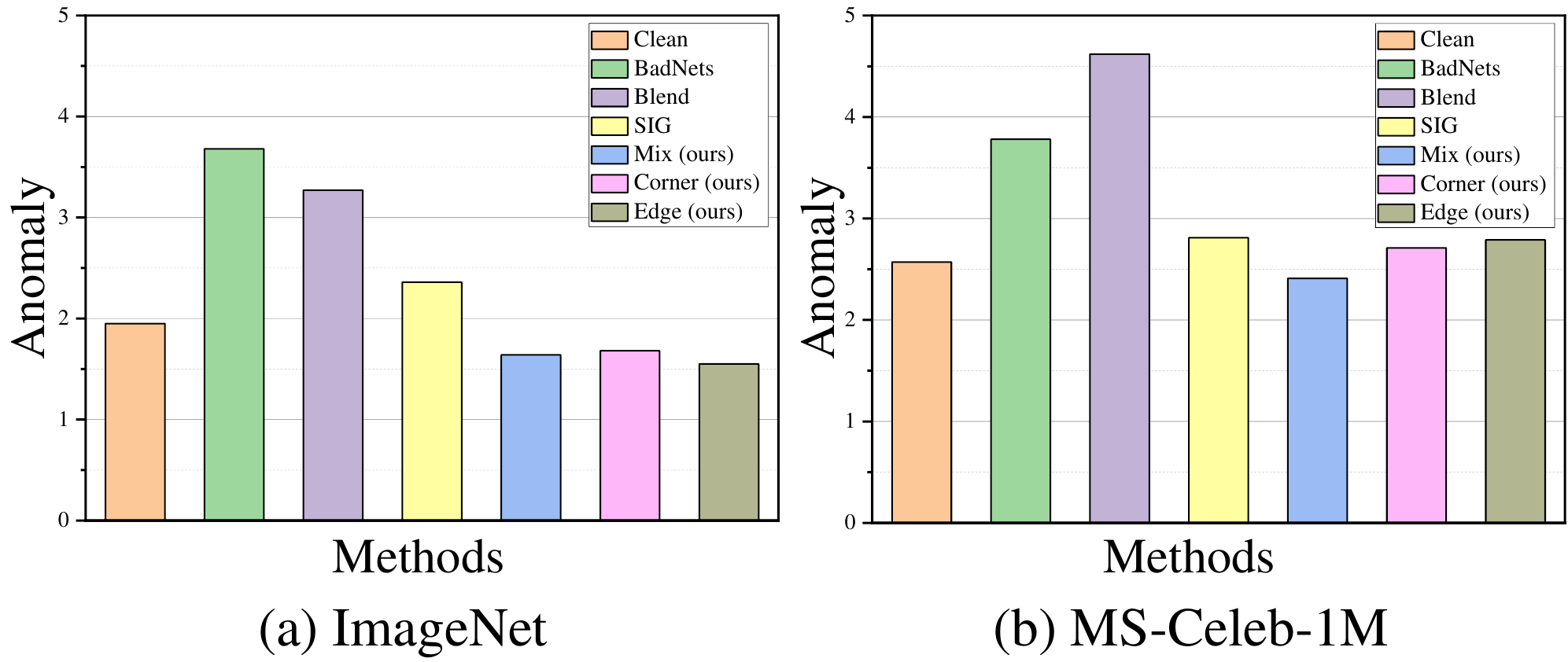}}
	\caption{The effect of different attack methods against Neural Cleanse defense method.}
	\label{figNC}
\end{figure}

\vspace{0.05cm}
\noindent{\bf Neural Cleanse.} Since the trigger corresponding to the backdoored model is much smaller than the trigger generated by the normal model, Neural Cleanse reversely constructs the triggers of each category and compares their anomaly index to determine whether the model is backdoored. We compare our method with BadNets, Blend, and SIG against Neural Cleanse defenses, which results are shown in Figure \ref{figNC}. As for the ImageNet dataset, the anomaly indexes of our feature injection modes are all lower than that of the clean model, and the anomaly indexes of BadNets, Blend, and SIG are all higher than that of the clean model.
As for the MS-Celeb-1M dataset, the anomaly index of the clean model has exceeded the threshold 2 set by Neural Cleanse, which is falsely recognized as backdoored. Despite the Neural Cleanse recognizing our models as backdoored, it should be noted that the anomaly index of the model trained on our poisoned data is close to that of the clean model. However, the anomaly indexes of BadNets and Blend are significantly higher than that of the clean model, indicating that our method is superior to BadNets and Blend.

\begin{figure}[t]
\setlength{\abovecaptionskip}{0.1cm}
	\centering
	\scalebox{0.9}{\includegraphics[width=\linewidth]{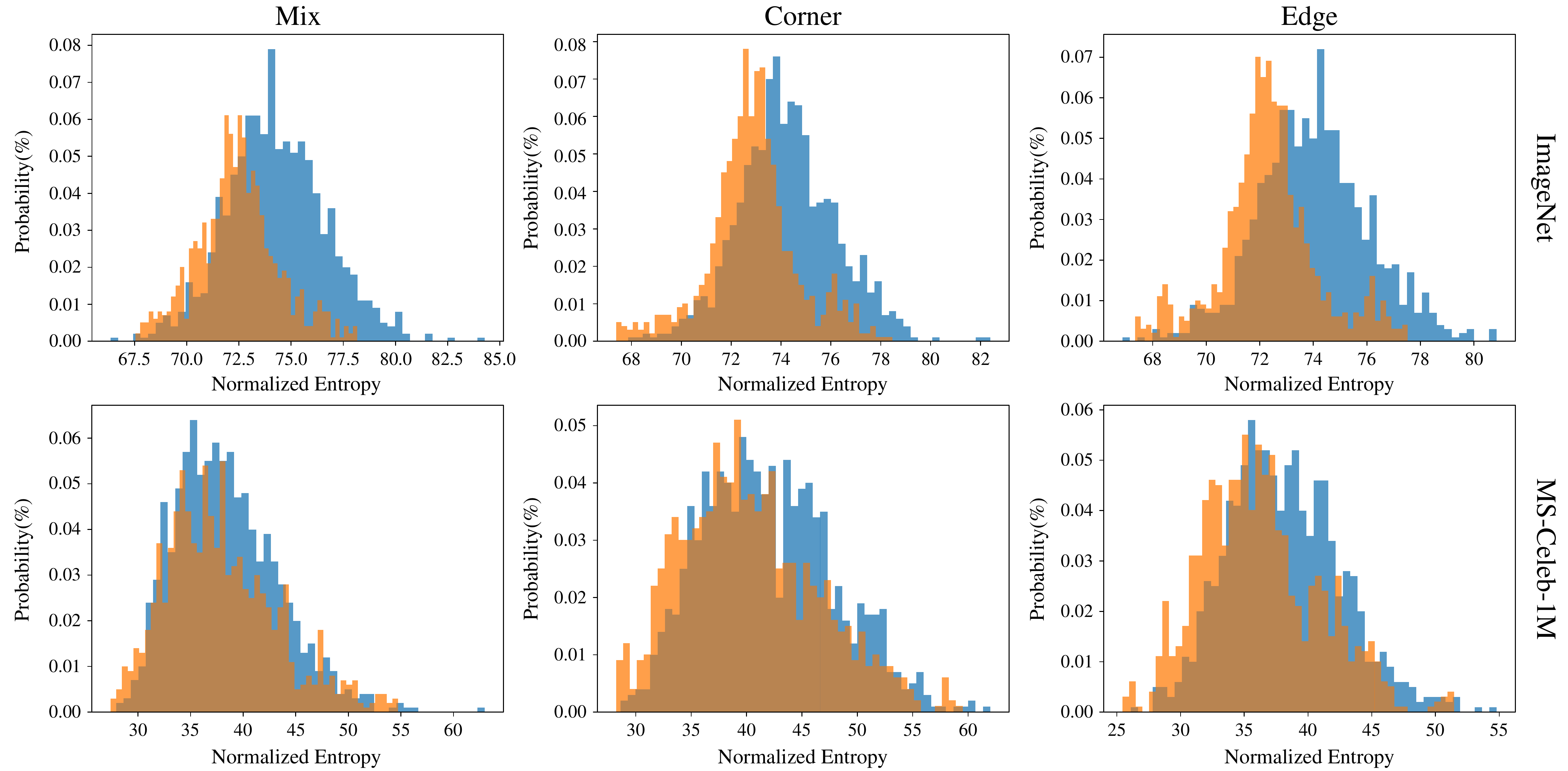}}
	\caption{Entropy distribution of clean and backdoor inputs for different attacks.}
	\label{figstrip}
\end{figure}

\vspace{0.05cm}
\noindent{\bf STRIP.} It utilizes the backdoor characteristic that any input containing a trigger will be classified as the target category. It mixes an input image with a set of clean images from different classes to inject strong perturbation and filters out trigger samples by comparing the information entropy of the model output classification result. As shown in Figure \ref{figstrip}, the entropy distribution of our attack model is very similar to that of the clean model. Thus STRIP fails to detect trigger samples of our attack.

\begin{figure}[t]
	\setlength{\abovecaptionskip}{0.1cm}
	\centering
    \scalebox{0.95}{
	\includegraphics[width=\linewidth]{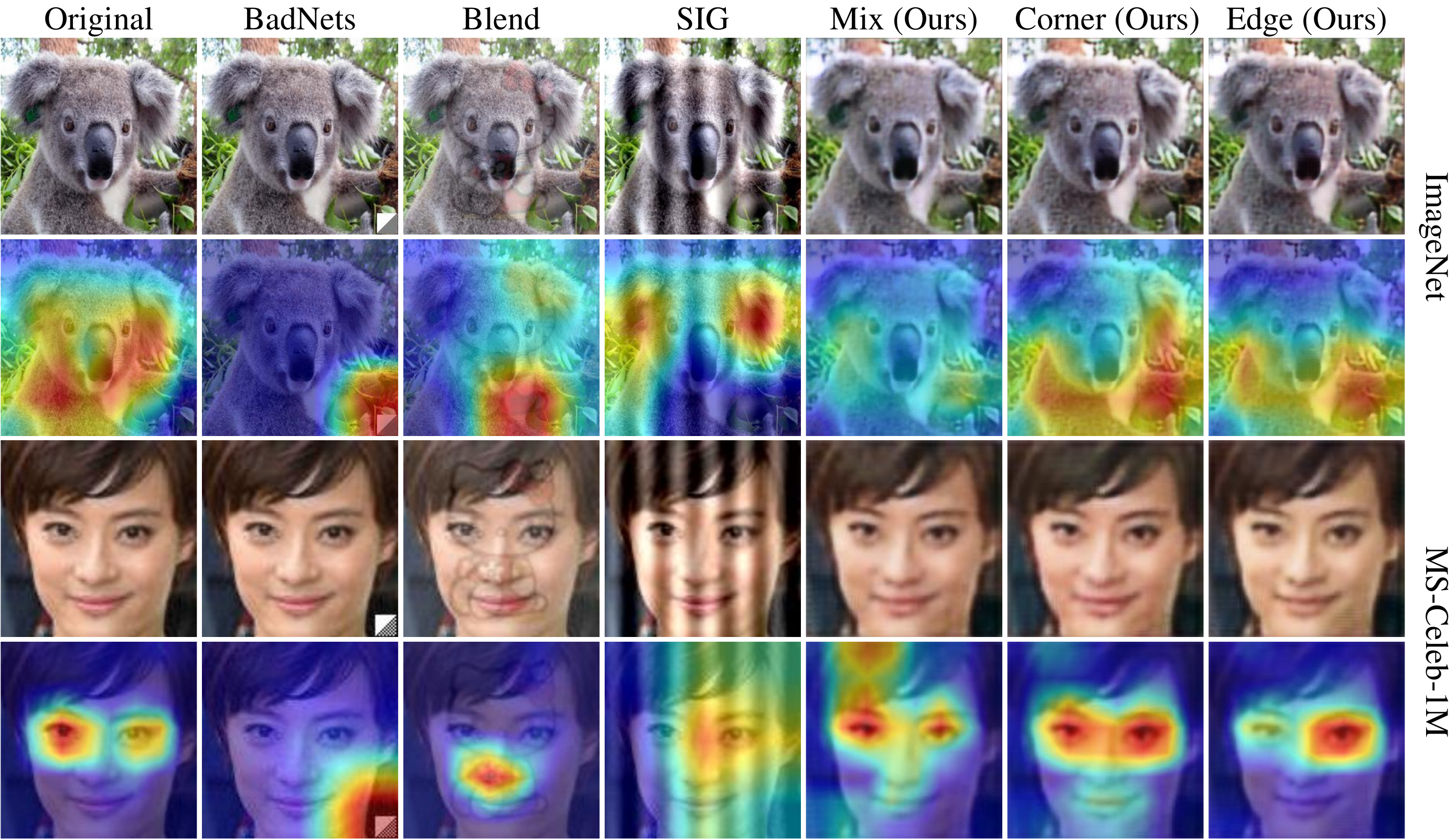}}
	\caption{The Grad-CAM of poisoned samples generated by different attacks.}
	\label{figcam}
\end{figure}

\vspace{0.05cm}
\noindent{\bf SentiNet.} SentiNet identifies trigger regions by the similarity of Grad-CAM of different samples. As shown in Figure \ref{figcam}, in both ImageNet and MS-Celeb-1M datasets, the Grad-CAM of the poisoned images produced by BadNets, Blend and SIG is significantly different from the clean images. While the Grad-CAM of the poisoned images generated by our three feature injection modes is similar to the clean images. Therefore, SentiNet is ineffective to our attacks.

\subsection{\bf Ablation Studies}\label{Ablation Studies}

\noindent{\bf Trigger Generator Transferability.}
In this part, we consider that in order to further reduce the attack overhead, the attacker uses the denoising autoencoder trained by dataset $D_A$ to generate poisoned dataset $D_B$ ($D_A$ and $D_B$ are for different classification tasks, and the image sizes of the two can be inconsistent).
The experimental results are shown in Tables \ref{transfer}, after the task transfer, our method basically has no effect on the CDA and ASR of the model, and has some effect on the concealment of poisoned images, but overall it is still better than BadNets, Blend and SIG.

\begin{table}[t]
\centering
\scalebox{0.7}{
    \begin{tabular}{c|cc|cc}
        \toprule %
        \multirow{2}*{\textbf{Methods}}&
        \multicolumn{2}{c|}{\textbf{\makecell[c]{MS-Celeb-1M for Encoder \\ImageNet for Classifer}}}&
        \multicolumn{2}{c}{\textbf{\makecell[c]{ImageNet for Encoder \\MS-Celeb-1M for Classifer}}}\\
        \cline{2-5}   %
        ~& \textbf{CDA} & \textbf{ASR} & \textbf{CDA} & \textbf{ASR} \\
        \midrule %
    Mix & 85.81 (99.48$\times$) & 99.98 & 98.25 (99.95$\times$) & 99.98 \\
    Corner & 85.76 (99.42$\times$) & 99.85 & 98.12 (99.82$\times$) & 99.97\\
    Edge & 85.71 (99.36$\times$) & 99.89 & 98.00 (99.69$\times$) & 99.98 \\
        \bottomrule %
    \end{tabular}}
    \caption{The effect of task transfer on model accuracy (CDA/ASR)}
    \label{transfer}
\end{table}

\begin{table}[!t]
\centering

  \scalebox{0.7}{
     \begin{tabular}{c|ccc|ccc}
        \toprule %
        \multirow{2}*{\textbf{Methods}}&
        \multicolumn{3}{c|}{\textbf{\makecell[c]{MS-Celeb-1M for Encoder \\ImageNet for Classifer}}}&
        \multicolumn{3}{c}{\textbf{\makecell[c]{ImageNet for Encoder \\MS-Celeb-1M for Classifer}}}\\
               \cline{2-7}   %
               ~& \textbf{MSE$\downarrow$} & \textbf{PSNR$\uparrow$} & \textbf{SSIM$\uparrow$} & \textbf{MSE$\downarrow$} & \textbf{PSNR$\uparrow$} & \textbf{SSIM$\uparrow$}  \\
        \midrule %
    Mix & 290.95 & 24.16 & 0.70 & 244.25 & 24.50 & 0.78 \\
    Corner& 321.22 & 23.85 & 0.70 & 205.03 & 25.30 & 0.80 \\
    Edge & 268.87 & 24.55 & 0.72 & 235.23 & 24.66 & 0.79\\
        \bottomrule %
        \end{tabular}}
    \caption{The effect of task transfer on trigger concealment}
    \label{transfer2}
\end{table}

\vspace{0.05cm}
\noindent{\bf Trigger Exclusivity.}
In this part, we measure whether the triggers generated by our method are exclusive, that is, whether a trigger generated based on image $x_1$ can still activate the backdoor after being added to a different image $x_2$. Ideally, sample-specific triggers should have high exclusivity.
For each testing image, we add a trigger generated from a randomly selected differing image and test its effectiveness. The results are shown in Tabel \ref{exclusive}, the trigger of our method is highly exclusive. Because the ASR is no more than 2\% among all cases.
\begin{table}[!t]
\centering
\scalebox{0.7}{
\begin{tabular}{c|c|c}
\toprule %
 \textbf{Feature Injection Mode}  &  \textbf{ImageNet} & \textbf{MS-Celeb-1M}  \\
\midrule %
Mix & 0.21 & 0.54 \\
Corner & 0.08 & 0.91 \\
Edge & 0.24 & 1.52\\
\bottomrule %
\end{tabular}}
\caption{ASR for models under inconsistent triggers (\%)}
\label{exclusive}
\end{table}

\vspace{0.05cm}
\noindent{\bf The Effect of Poisoning Ratio $\rho$.} In previous experiments, we set the poisoning ratio $\rho$ to 10\%. We now compare the effects of different $\rho$ on the CDA and ASR of the infected model. As shown in Table \ref{tablePoisoningRatio}, the reduction of $\rho$ hardly affects the CDA of the model, because the poisoned data only occupies a small part of the training set, and the CDA is mainly affected by the dominant clean dataset.
The reduction of $\rho$ mainly affects the ASR of the model. When it is reduced to 1\%, the degradation of ASR on the ImageNet dataset is about 6\%, while the loss on the MS-Celeb-1M dataset is negligible that is only 0.5\%. Overall, our method performs well in terms of robustness even with a small poisoning ratio.

\begin{table}[!t]
\renewcommand\arraystretch{1.2}
\centering
\scalebox{0.58}{
    \begin{tabular}{c|ccc|ccc}
        \toprule %
        \multirow{2}*{\textbf{ \makecell[c]{Poisoning\\ratio} }}&
        \multicolumn{3}{c|}{\textbf{ImageNet}}&
        \multicolumn{3}{c}{\textbf{MS-Celeb-1M}}\\
               \cline{2-7}   %
               ~& \textbf{Mix} & \textbf{Corner} & \textbf{Edge} & \textbf{Mix} & \textbf{Corner} & \textbf{Edge} \\
        \midrule %
    1\% & 85.71/93.74 & 85.65/95.08 & 85.88/94.63 & 97.75/99.84 & 97.94/99.58 & 97.91/99.53\\
    5\% & 85.84/99.64 & 85.73/99.67 & 85.82/99.74 & 98.01/99.90 & 98.05/99.94 & 97.88/99.99\\
        \bottomrule %
    \end{tabular}}
    \caption{The effect of different poisoning ratios on the backdoored model (CDA/ASR)}
    \label{tablePoisoningRatio}
\end{table}

\section{Conclusion}\label{Conclusion}

This work constructively incorporates denoising autoencoder with dedicated feature inject mode selection to create imperceptible sample-specific triggers for backdoor attacks.
The denoising autoencoder denoises images injected with a specific (visible) feature to generate trigger images that are almost visually the same as the original clean image content.
This method does not require the attacker to know the structure of the victim model, and its  trigger generator owns preferable task-transferability. Comprehensive experiments have validated the effectiveness of the method without affecting the accuracy of the model while achieving a high attack success rate, and high exclusivity. Moreover, it has successfully evaded four diverse mainstream defenses.

\bibliographystyle{named}
\bibliography{reference}

\end{document}